\documentclass[11pt]{article}
\usepackage{moriond,epsfig}

\def\Journal#1#2#3#4{{#1} {\bf #2}, #3 (#4)}

\def\NIMA{\em {Nucl. Instrum. Methods} A}

\def\PRD{\em {Phys. Rev.} D}
\def\PRB{\em {Phys. Rev.} B}

\begin{document}
\vspace*{4cm}
\title{STATUS AND OUTLOOK OF THE EDELWEISS WIMP SEARCH}

\author{ M. LUCA\\
FOR THE EDELWEISS COLLABORATION }

\address{Institut de Physique Nucl\'eaire de Lyon, Universit\'e Claude Bernard Lyon 1\\
4 rue Enrico Fermi, 69622 Villeurbanne Cedex, France\\
E-mail : luca@ipnl.in2p3.fr}

\maketitle\abstracts{
EDELWEISS is a direct dark matter search using cryogenic germanium heat--ionisation detectors, located in the Modane underground
laboratory beneath the Alps. We summarize the final results of EDELWEISS I, which deployed up to almost one kg of detectors in its final 
stage. EDELWEISS II recently started commissioning runs. With an increased detection mass and better shielding, this stage aims to gain two orders of magnitude
in sensitivity and to serve
as a testbed for a larger, ton--scale experiment.}

\section{Introduction}

Understanding the nature of dark matter in the universe is a major challenge for modern cosmology and astrophysics. One of the
well--motivated candidates is the particle generically named WIMP (Weakly Interacting Massive Particle), such as the lightest
supersymmetric particle. The main constraints when attempting to detect WIMPs are low event rate (less than a few events per kg of
detector and per year) and small recoil energies (a few keV).

EDELWEISS (Exp\'erience pour D\'etecter les WIMPs en Site Souterrain) is a direct dark matter detection experiment using the elastic
scattering of WIMPs off target nuclei. EDELWEISS is situated in the Modane Underground Laboratory, in the Fr\'ejus highway tunnel between
 France and Italy. An overburden of 1700 m of rock, equivalent to 4800 m of water, reduces the muon flux down to \mbox{4.5 $\mu$/m$^2$/day}, that is about 10$^6$
 times less than at the
 surface.

\section{EDELWEISS I final results}

\begin{figure}[!tb]

% to be included top second page

\begin{center}
\includegraphics[width=6.5cm]{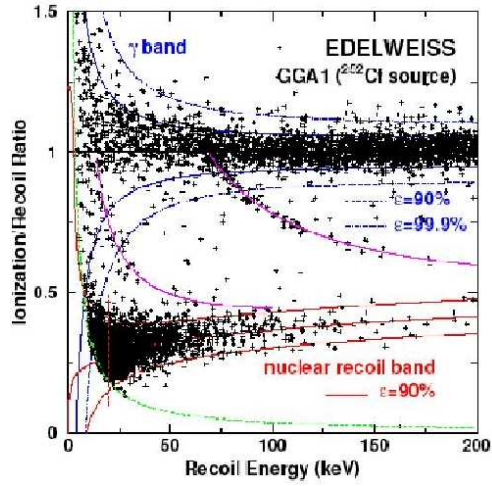}
\caption{Ratio between the ionisation and the recoil energy versus the recoil energy. The separation between the electronic
 and the nuclear recoil band is excellent down to an energy of about 15 keV.}
\label{fig:biplot}
\end{center}
\end{figure}

EDELWEISS uses cryogenic germanium detectors, in a dilution fridge working at about 17 mK. Each detector has a NTD--Ge thermistor
 that measures the heat signal and two Al electrodes for the ionisation.~\cite{ab} This technique of measuring two signals 
simultaneously allows an event by event discrimination between
electronic (induced by photons or electrons) and nuclear (induced by neutrons or WIMPs) recoils
(Fig.~\ref{fig:biplot}).~\cite{ps}

\begin{figure}[!tb]

%to be included top third page

\begin{center}
\includegraphics[width=6.5cm]{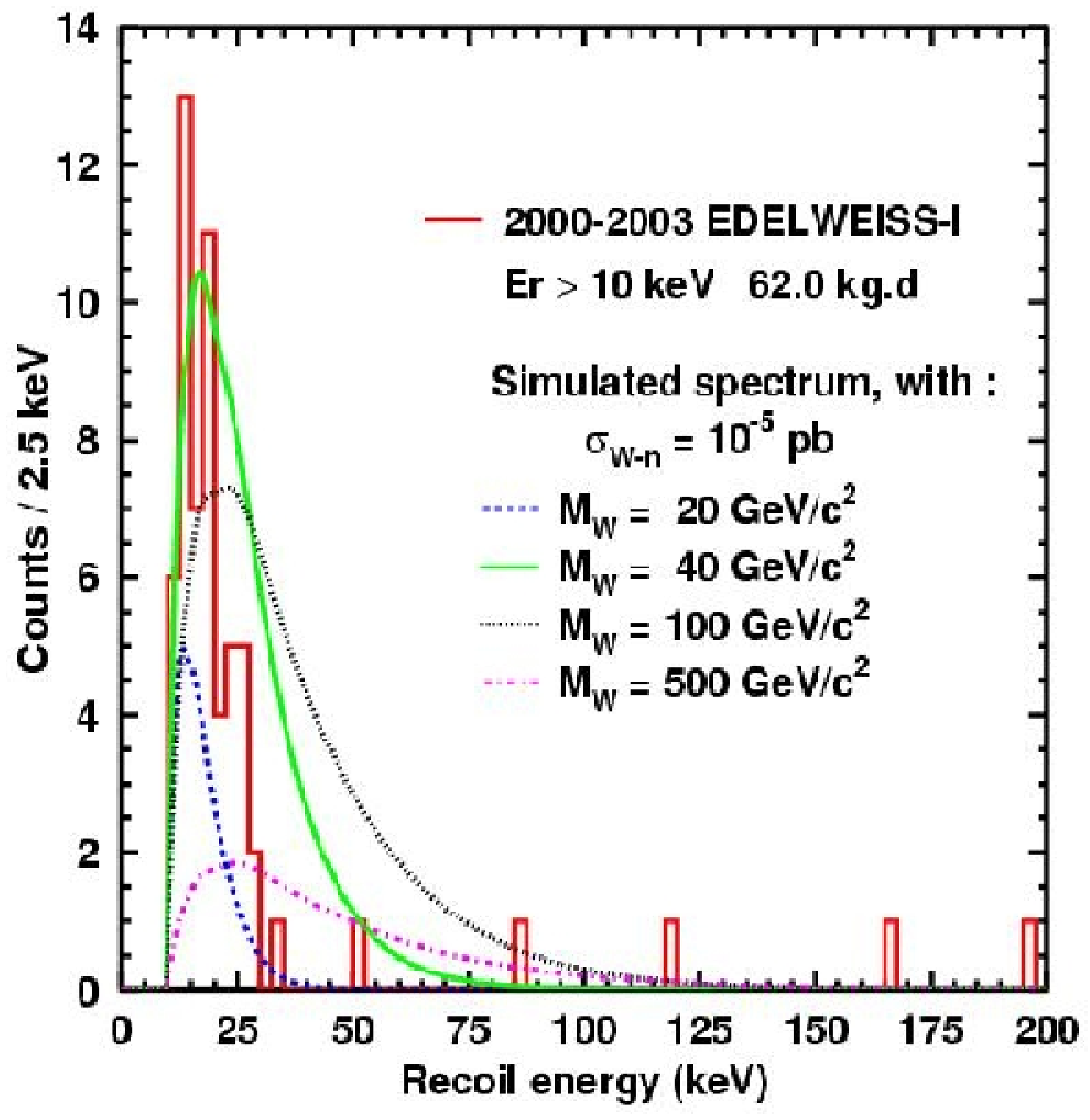}
\caption{Energy spectrum of the EDELWEISS I experimental data compared to simulated spectra for different WIMP masses in the range of
interest.}
\label{fig:spectrum}
\end{center}
\end{figure}

EDELWEISS I used 320 g Ge detectors during several campaigns. Between 2002 and 2003, three 320 g detectors have been operated simultaneously
in a cryostat shielded by 10 cm of Cu, 15 cm of Pb, 7 cm of internal roman Pb and 30 cm of paraffin.  After a total fiducial exposure of 62
kg$\cdot$day with an effective threshold of 15 keV, 59 events have been observed in the nuclear--recoil band.~\cite{vs} As shown in Fig.~\ref{fig:spectrum}, most of the events are
at low energy, between 10 and 30 keV. The simulated spectra of WIMPs having a scattering cross section on nucleons of 10$^{-5}$ pb and
masses of 20, 40, 100 and 500 GeV/c$^2$ show that the events in the nuclear recoil band can not be explained by a single WIMP mass, a
part of the spectrum has to be attributed to a non--WIMP background.

\begin{figure}[!tb]

%to be included second page, beneath the first figure

\begin{center}
\includegraphics[width=6.5cm]{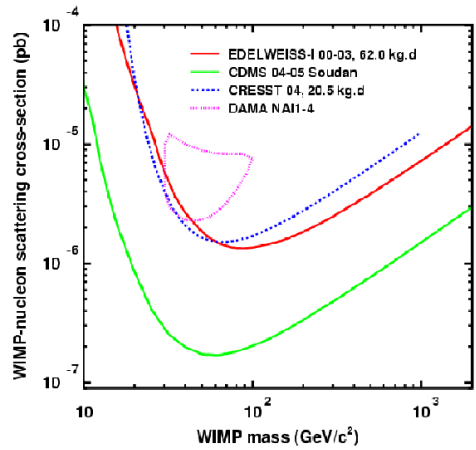}
\caption{Spin independent exclusion limits on WIMPs, obtained with standard astrophysical assumptions within a 90\% C.L.. Solid dark curve
is
the EDELWEISS I final limit, dashed curve is the CRESST limit from CaWO$_4$~\protect\cite{ga}, solid light curve is the CDMS limit from the Soudan
mine~\protect\cite{da} and closed contour is the allowed region at 3$\sigma$ C.L. from DAMA NaI1-4 annual modulation data~\protect\cite{rb}.}

\label{fig:exclusion}
\end{center}
\end{figure}

The two main sources of background that can mimic WIMP events are the
 mis--collected ionisation events and the neutrons. A two detector coincident event with both hits in the nuclear recoil band has
  been observed. The most likely source of this event is a neutron scattering.
 Monte--Carlo simulations predicted about two neutron single events for a total exposure of 62 kg$\cdot$day. Nevertheless, the exclusion 
 limit in Fig.~\ref{fig:exclusion} is derived without any background subtraction.  It confirms, after a longer fiducial exposure, 
 the previously published EDELWEISS I limits~\cite{be}.

EDELWEISS I, once the most sensitive direct dark matter search, became limited mainly because of the radioactive background (neutrons and surface
events) and detection mass (the cryostat could not host more than three 320 g detectors). Therefore, the experiment has been
stopped in March 2004, and is now replaced by EDELWEISS II.

 \section{EDELWEISS II}
 
The second stage of the experiment is EDELWEISS II. By diminishing the radioactive background and increasing the detection mass, it should
 gain a factor of 100 in sensitivity compared to EDELWEISS I. Among the numerous improvements in EDELWEISS II, the main ones are: a new,
larger, low consumption cryostat, a larger detection mass, improved detectors, new shieldings and an active muon veto, a class 100 clean room,
new electronics and acquisition system.

In order to host up to 120 detectors, the new cryostat is larger (50 l) and has an innovative reversed geometry. Three pulse tubes replace
liquid nitrogen cooling and a cold vapor reliquefier reduces helium consumption. The compact and hexagonal arrangement of the detectors
should increase the neutron coincidence rate.

\subsection{Radioactive background}

The main limiting factor for EDELWEISS I was the radioactive background.

In a Ge detector, the ionisation signal coming from particles that interact very close to the surface can be mis--collected,
hence the resulting events appearing in the nuclear recoil band. One way of dealing with this problem
is by depositing a 60 nm Ge or Si amorphous layer on the crystal surface which diminishes the number of surface events~\cite{ts}. All
detectors used for EDELWEISS I 2002--2003 runs had such layers.
 Concentric electrodes provide a radial sensitivity allowing to select events occuring in the central part of the detector~\cite{om} where the
 electric field is more homogeneous and the detector better shielded from its environement.

 A promising R\&D project on surface events uses the sensitivity of NbSi thin film thermometers to athermal phonons~\cite{sm}. As 
 surface
 events have a higher athermal component, they can be identified and excluded during analysis. A 200 g detector with NbSi thin films on
 each end has already been tested in the EDELWEISS I cryostat with good results~\cite{sm} and is part of the EDELWEISS II setup for the 
 first commissioning runs.

 High energy neutrons are hard to moderate and can penetrate the shielding. When interacting with the detector, neutrons can mimic WIMP events,
 therefore they are an important issue for EDELWEISS II. Some of the ways of dealing with this problem in EDELWEISS II
  are  a better shielding, 
 a muon veto and the increase in the number of neutron coincident events.
 
 \subsection{SciCryo}

 Another R\&D project, \emph{SciCryo}\footnote{Institut de Physique Nucl\'eaire de Lyon (IPNL), Institut d'Astrophysique Spatiale Orsay
 (IAS), 
 Max Planck Physik Munich (MPP), Laboratoire de Physico--Chimie des Materiaux Luminescents Lyon (LPCML)}, studies the
 possibility of using heat--scintillation bolometers for dark matter detection. For now, \emph{SciCryo} has focused on light
 targets, complementary to Ge detectors 
 and interesting
 for both neutron detection and spin--dependent interactions. So far, the most promising light target is Al$_2$O$_3$. Several nominally pure
  sapphire crystals have been tested in
 Lyon at room temperature and all of them have been seen to scintillate. Some of the crystals have also shown very encouraging low 
 temperature
 light yields in tests performed at the IAS and the MPP. As a result of these tests, a 50 g IAS sapphire heat--scintillation detector has been
 included in the first EDELWEISS II commissioning runs to check its compatibility with the setup.

 \subsection{EDELWEISS II near future}
 
 Since January 2006, commissioning runs are taking place in order to check the level of microphonics and to test the new electronics and
 acquisition system. For now, eight detectors have been mounted in the cryostat: 4 Ge/NTD EDELWEISS I detectors, 2 new 320 g Ge/NTD
 detectors and 2 R\&D detectors (a 200 g Ge/NbSi and a 50 g sapphire heat--scintillation detector). The goals for this year are to
 increase the number of bolometers to 28  and to define the next stage that may include up to 120 detectors.

\section{Conclusion}

EDELWEISS I, once the most sensitive direct dark matter search has been stopped in 2004. An important result of EDELWEISS I was the
identification of the two main sources of background, surface events and neutrons. EDELWEISS II should
gain a factor of 100 in sensitivity thanks to improved detectors, shielding, cleanliness and a higher detection mass.
 The main R\&D projects concern the NTD and NbSi
Ge detectors as well as the cryogenic heat--scintillation detectors. Preliminary results are expected in 2006.

\section*{Acknowledgments}

This work has been partially funded by the EEC Network program under the contract HPRN--CT--2002--00322, the ILIAS integrating activity
contract RII3-CT-2003-506222 and by 
Agence Nationale de la Recherche grant JC05\_41907.

\newpage

\section*{References}
\begin{thebibliography}{99}




\bibitem{ab}A. Benoit {\it et al.}, \Journal{\PRB}{479}{8}{2000}.

\bibitem{ps}P. Di Stefano {\it et al.}, \Journal{Astropart. Phys.}{14}{329}{2001}.

\bibitem{vs}V. Sanglard {\it et al.}, \Journal{\PRD}{71}{122002}{2005}.

\bibitem{be}A. Benoit {\it et al.}, \Journal{\PRB}{545}{43}{2002}.

\bibitem{ga}G. Angloher {\it et al.}, \Journal{Astropart. Phys.}{23}{325}{2005}.

\bibitem{da}D. Abrahams {\it et al.}, \Journal{\PRD}{66}{122003}{2002}.

\bibitem{rb}R. Bernabei {\it et al.}, \Journal{\PRB}{480}{23}{2000}.

\bibitem{ts}T. Shutt {\it et al.}, \Journal{\NIMA}{444}{340}{2000}.

\bibitem{om}O. Martineau {\it et al.}, \Journal{\NIMA}{530}{426}{2004}.

\bibitem{sm}S. Marnieros {\it et al.}, \Journal{\NIMA}{520}{185}{2004}.

\end{thebibliography}
\end{document}